\documentclass{article}
\usepackage{spconf,amsmath,graphicx}
\usepackage{amssymb}
\usepackage{multirow, multicol}
\usepackage{hyperref}
\usepackage{footmisc}

\title{Self-Supervised Learning for Few-Shot Bird Sound Classification}
\twoauthors
    {Ilyass Moummad, Nicolas Farrugia}
    {IMT Atlantique, Lab-STICC, \\UMR CNRS 6285, Brest, France,}
    {Romain Serizel}
    {University of Lorraine, CNRS, Inria, \\Loria, 54000, Nancy, France,}
\begin{document}
%
\maketitle
\begin{abstract}
Self-supervised learning (SSL) in audio holds significant potential across various domains, particularly in situations where abundant, unlabeled data is readily available at no cost. This is pertinent in bioacoustics, where biologists routinely collect extensive sound datasets from the natural environment. In this study, we demonstrate that SSL is capable of acquiring meaningful representations of bird sounds from audio recordings without the need for annotations. Our experiments showcase that these learned representations exhibit the capacity to generalize to new bird species in few-shot learning (FSL) scenarios. Additionally, we show that selecting windows with high bird activation for self-supervised learning, using a pretrained audio neural network, significantly enhances the quality of the learned representations.\footnote{Our code is released at:}\let\thefootnote\relax\footnotetext{\url{https:/github.com/ilyassmoummad/ssl4birdsounds}}
\end{abstract}
\let\thefootnote\relax\footnotetext{This work is co-funded by AI@IMT program and OSO-AI company}
%
\section{Introduction}
\label{sec:intro}

Bioacoustics, the study of sounds produced by living organisms, provides valuable insights into species diversity, behavior, and ecosystem health. Bird activity, in particular, can reflect changes in habitats and climate~\cite{birdnet}. Deep learning, capable of automatically learning patterns from large datasets, emerges as a powerful tool for the automatic identification of bird species based on their unique acoustic signatures~\cite{dan2022}.

In the pursuit of leveraging technological advancements, BirdCLEF emerges as an annual competition dedicated to identifying endangered bird species through bioacoustics~\cite{birdclef2020}. The data for BirdCLEF is sourced from the collection of bird sounds Xeno-Canto (\textit{https://xeno-canto.org/}). BirdCLEF aims to enhance techniques in bioacoustics to tackle the challenges associated with the identification of bird species by providing labeled datasets of bird sounds. In this context, BirdNet~\cite{birdnet} emerges as a model specifically trained on thousands of hours of Xeno-Canto bird data using supervised learning (SL). 
However, the requirement for abundant labeled training data in SL hinders progress, particularly for species with limited documentation~\cite{nolasco}. Nolasco et al.~\cite{nolasco} address this problem by reformulating bioacoustic sound event detection as a collection of small tasks, each having few labeled data per animal sound, aligning with the FSL framework. 

With the goal of advancing FSL for bioacoustic, Ghani et al.~\cite{ghani} show that embeddings derived from large-scale audio models, particularly BirdNet and Google Perch, trained on bird sounds using SL, exhibit strong performance in bioacoustic few-shot classification tasks. Considering the broader scope of FSL for audio classification, MetaAudio~\cite{metaaudio} proposes a benchmark of few-shot classification, covering a variety of domains including bird sounds. The authors split the training set of BirdCLEF~2020~\cite{birdclef2020}, into three distinct sets of classes, representing new training, validation and test sets. The goal is to transfer knowledge from a set of classes, addressing the challenge of $n$-way $k$-shot sound classification.

This study explores SSL of bird sound representations on the MetaAudio split of BirdCLEF. Challenges arise in defining a pretext task to learn representations of bird sounds. We investigate three methods: SimCLR~\cite{simclr}, Barlow Twins~\cite{bt}, and FroSSL~\cite{frossl}. Focusing on pretext-invariant representation learning, we exclude reconstruction methods due to their requirements for large datasets, and their demonstrated poor performance in low-shot classification~\cite{msn}.

Data augmentation plays a crucial role in the SSL of invariant representations~\cite{simclr}. The specific invariance that a model needs to learn for efficient representations in bioacoustics is not fully understood~\cite{dan2022}. In this context, we operate under the assumption that domain-agnostic data augmentation, which does not assume any specific strong invariance on bird sounds, should prove effective for this task.

Beyond this aspect, in practical scenarios with lenghty audio files, particularly in bioacoustics, selecting training segments becomes crucial for the encoder's quality, as choosing segments with only background noise or silence will be detrimental to the performance. Denton~et~al.~\cite{denton} apply a wavelet peak detector to find frames with high energy. A recent foundation model in computer vision, DINOv2, selects training images using a pretrained model on a large uncurated dataset to retrieve similar images in a curated dataset~\cite{dinov2}. Similarly, we use a pretrained audio neural network (PANN) on AudioSet~\cite{panns}, a dataset for general-purpose audio tagging~\cite{audioset}, to select the segment with highest bird class activation.

\section{Method}
\label{sec:method}
In this section, we describe one SL and three SSL methods for pre-training representations of bird sounds. Subsequently, we describe the few-shot classification task for evaluating the learned representations.
\subsection{Representation Learning}
\label{ssec:ssl}

We describe three SSL methods: SimCLR~\cite{simclr} (sample contrastive), Barlow Twins~\cite{bt} (dimension contrastive), and FroSSL~\cite{frossl} (both sample and dimension contrastive), along with SupCon~\cite{supcon} as a supervised method for reference. Let ${X}$ be a batch of $N$ examples, ${X_1}$ and ${X_2}$ represent the two resulting batches obtained by applying two transformations to each example in $X$. 
Let ${Z_1}$ and ${Z_2}$ be the two batches of embeddings extracted from ${X_1}$ and ${X_2}$, respectively.

\subsubsection{SimCLR}
\label{sssec:simclr}

Let $\boldsymbol{z}_{i}$ and $\boldsymbol{z}_{j}$ be the $\ell^2$-normalized embeddings of two views $\boldsymbol{x}_{i}$ and $\boldsymbol{x}_{j}$ coming from $X_1$ and $X_2$, respectively.
The loss function for SimCLR~\cite{simclr} computed for two similar views $(i,j)$ originating from the same original example as:
\begin{equation}
    \label{simclr}
    \mathcal{L}^{SimCLR}_{(i,j)} = -\log\frac{\textit{exp}\left(\boldsymbol{z}_i\boldsymbol{\cdot}\boldsymbol{z}_{j}/\tau\right)}{\sum\limits_{k\neq i}\textit{exp}\left(\boldsymbol{z}_i\boldsymbol{\cdot}\boldsymbol{z}_{k}/\tau\right)},
\end{equation}
where $\tau\in\mathbb{R}^{+*}$ is a scalar temperature parameter and $\boldsymbol{z}_{k}\in \text{concat}([Z_1, Z_2])$.
The final loss SimCLR for the augmented batch is written as: 
\begin{equation}
    \label{SIMCLR}
    \mathcal{L}^{SimCLR} = \sum_{i,j}\frac{1}{2}(\mathcal{L}^{SimCLR}_{(i,j)}+\mathcal{L}^{SimCLR}_{(j,i)}),
\end{equation}

SimCLR pulls in the embedding space similar views while pushing appart dissimilar ones.
\subsubsection{Barlow Twins}
\label{sssec:bt}

Let ${C}\in\mathbb{R}^{N\times D}$ (with $D$ the feature space dimension) be the cross-correlation matrix computed between ${Z_1}$ and ${Z_2}$ (assumed to be mean-centered along the batch dimension) computed as: ${C}=\frac{1}{N}{Z_1}^\mathsf{t}\cdot{Z_2}$ where $^{\mathsf{t}}$ is the transpose operator and $\cdot$ is the matrix multiplication.
The loss function for Barlow Twins (BT)~\cite{bt} for the batch of embeddings ${Z_1}$ and ${Z_2}$ with cross-correlation matrix ${C}$ is computed as:

\begin{equation}
    \label{BT}
    \mathcal{L}^{BT} = \sum_i (1-C_{ii})^{2} + \lambda \sum_i\sum_{j\neq i} C_{ij}^{2},
\end{equation}
BT causes the embedding vectors of two transformed versions of an example to be similar, while minimizing the redundancy between the components of these vectors.

\subsubsection{FroSSL}
\label{sssec:fro}
${Z_1}$ and ${Z_2}$ are mean-centered along the batch dimension and normalized by their respective Frobenius norms.
The Frobenius norm of a matrix ${Z}\in\mathbb{R}^{N\times D}$ is defined as: $\|Z\|_{F}^{2}=\sum_{i=1}^{N}\sum_{j=1}^{D}Z_{ij}^{2}=\sum_{k}^{min(N,D)}$.
The loss function for FroSSL is defined as:
\begin{equation}
    \label{FRO}
    \mathcal{L}^{Fro} = \textit{MSE}(Z_1,Z_2) + \lambda\,\textit{R}(Z_1,Z_2),
\end{equation}
where:
\begin{equation}
    \label{MSE}
    \textit{MSE}(Z_1,Z_2) = \frac{1}{N}\sum_{i=1}^{N}\|Z_{1}^{i}-Z_{2}^{i}\|_{2}^{2},
\end{equation}
and:
\begin{equation}
    \label{REG}
    \textit{R}(Z_1,Z_2) = \log(\|Z_{1}^{\mathsf{t}}\cdot Z_1\|_{F}^{2}) + \log(\|Z_{2}^{\mathsf{t}}\cdot Z_2\|_{F}^{2}),
\end{equation}
FroSSL minimizes the Euclidean distance between the embedding vectors of two transformed versions of an example, and the Frobenius norms of their respective covariances. Given the duality of Frobenius norm: $\|Z_{i}^{\mathsf{t}}\cdot Z_i\|_{F}^{2}=\|Z_{i}\cdot Z_{i}^{\mathsf{t}}\|_{F}^{2}$, FroSSL can be considered both sample and dimension contrastive. We refer the reader to the original work for more theoretical details~\cite{frossl}.
In the original formulation, the authors did not introduce the hyperparameter tradeoff $\lambda$ between the invariance term $\textit{MSE}$ and the variance term $\textit{R}$. Their intuition and experiments suggested that the optimal tradeoff occurred when the two terms were equally weighted. However, we found in our exploration that the coefficient of the variance term has to be small for FroSSL to converge.
\subsubsection{SupCon}
\label{sssec:supcon}

SupCon~\cite{supcon} is the supervised extension of SimCLR where the two views are two examples sharing the same label. Using the same notations as Eq~\ref{simclr}, the loss of SupCon is defined as:
\begin{equation}
    \label{scl}
    \mathcal{L}^{SupCon} = \sum_{i}\frac{-1}{|P(i)|}\sum_{j\in P(i)}\log{\frac{\text{exp}\left(\boldsymbol{z}_i\boldsymbol{\cdot}\boldsymbol{z}_j/\tau\right)}{\sum\limits_{k\neq i}\text{exp}\left(\boldsymbol{z}_i\boldsymbol{\cdot}\boldsymbol{z}_k/\tau\right)}},
\end{equation}
where ${P(i)={\{j :{{y}}_j={{y}}_i}\}}$ is the set of indices of all views sharing the same label with $i$. $|P(i)|$ is its cardinality.

\subsection{Few-Shot Learning}
\label{ssec:ssl}

We consider FSL tasks sampled from the test split. A $n$-way $k$-shot few-shot task consists of a support set of $k$ labeled examples from $n$ classes to learn on, and a query set of test examples. Following our previous work~\cite{moummad}, we adopt a simple inference method known as the nearest prototype classifier. This approach involves computing Euclidean distances to make predictions. Let ${\Bar{{z}_i}}$ be the prototype for each class label $i$ from the support set, for each query from the query set of feature vector $q$, we predict its label $i_q$ as:
\begin{equation}
    \label{prediction}
    i_q= \arg\min_i \|q-\Bar{{z}_i}\|_2,
\end{equation}
\section{Experiments}
\label{sec:exp}

In this section, we describe the dataset, the strategy for selecting windows, data augmentations, and implementation details.

\subsection{Data}
\label{sec:dataset}

In our study, we conduct experiments on BirdCLEF2020 (Pruned) dataset from MetaAudio~\cite{metaaudio}. This dataset is derived from the training set of BirdCLEF2020 by removing samples longer than 3 minutes and classes with less than 50 samples. It comprises 44,543/6,222/12,600 examples across 501/72/142 classes for the train, validation, and test sets, respectively.
\begin{table}[h]
\caption{Data distribution in BirdCLEF 2020.}
\label{Tab:data}
\begin{center}
\resizebox{\columnwidth}{!}{\begin{tabular}{lcccc}
\hline
Dataset               & N°Classes & N°Samples & Duration \\ \hline
BirdCLEF~2020          & 960       & 72,305    & 3s-30m   \\ \hline
BirdCLEF~2020~(Pruned) & 715       & 63,365    & 3s-3m  \\ \hline
\end{tabular}}
\end{center}
\end{table}

\subsubsection{Data Selection}
\label{sec:ds}

In our approach to handling audio signals of varying durations during training, we pad short signals to a fixed duration using a circular padding. For long audio signals, we employ either temporal proximity or PANN selection (PS). Temporal proximity randomly selects two nearby segments, encouraging the model to exploit temporal consistency~\cite{jansen}. PS involves dividing the audio signal into chunks of a fixed duration, passing them through a PANN on AudioSet, and selecting the chunk with the highest activation for the bird class. It's worth mentioning that PANN undergoes supervised training to classify over 500 diverse classes, among which the bird class constitutes approximately 1\% of the dataset. 

\subsubsection{Data Augmentation}
\label{sec:da}

We explore a variety of data augmentations drawn from the literature on SSL for audio~\cite{jansen,uclser}. Through ablation experiments, not shown in this work, we observe that certain techniques, such as frequency shift, additive white Gaussian noise, compression, and random resized crop, are detrimental to learning representations for bird sounds.

Loris et al.~\cite{bioacda} explore three data augmentation families for animal sounds within a supervised classification framework, comparing spectrogram augmentations to image and signal augmentations. While their experiments demonstrate the superiority of spectrogram augmentations, the absence of an ablation study assessing the individual impact of each technique raises uncertainties about their suitability for SSL. It is important to note that label information in supervised settings may mitigate negative effects of inappropriate data augmentation, a condition not applicable to self-supervised invariant learning (Table~\ref{Tab:abla}). Therefore, we present a simple domain-agnostic data augmentation pipeline without specific hypothesis on spectrotemporal properties of bird sounds. It incorporates time shift~\cite{uclser} of spectrograms along the time axis to encourage the model to learn features invariant to temporal shifts. It also includes spectrogram mixing~\cite{uclser, jansen}, a process of random interpolation of spectrograms. This generates noisy views to encourage the model to learn features robust to various forms of noise. Finally, the pipeline includes the random masking of consecutive frequency bands and time steps using SpecAugment~\cite{specaug, uclser}, guiding the model to focus on capturing higher-level and more abstract representations.
\subsection{Implementation Details}
\label{sec:imp}

Following MetaAudio~\cite{metaaudio}, we set the duration of training examples to 5 seconds. Signals are sampled at 16~kHz. We compute mel spectrogram features from FFT of size 1024, a hop length of 320, and a number of mels of 128. In order to generate two views, we use one of the two strategies: Temporal proximity, where we select two windows with a minimum of 60\% overlap; or PS, where we choose a single window using CNN14, a PANN on Audioset. We then apply data augmentation to the resulting windows.
Time shift is circular such that elements shifted beyond the last timestep are reintroduced at the first timestep. Spectrogram mixing linearly interpolates samples with random elements in the batch using a coefficient sampled uniformly between 0.6 and 1. SpecAugment masks three consecutive blocks of 30 frequency bands and 10 timesteps. The hyperparameters for data augmentation were tuned on the validation split.

For all our experiments, we train MobileNetV3-Large~\cite{mobilenetv3} from scratch using AdamW optimizer with a batch size of 256, a learning rate of 1e-3, and a weight decay of 1e-6 for 100 epochs. For BT and FroSSL, we employ a regularization coefficient of 1e-2. In the case of SimCLR and its supervised counterpart SupCon, a temperature of 1 is utilized.
For inference, when PS is employed, the chosen segment with PANN serves as the prototype. Otherwise, we divide long test audio files into 5s chunks, average their feature vectors to get one prototype (for both query and test audio files). The features are then centered using the mean of the support set prototypes and are $\ell^2$-normalized before computing the Euclidean distance to make the predictions. We conduct 10,000 5-way 1-shot test tasks for each run, and all our experiments represent the average results obtained from three runs.

\section{Results and Discussion}
\label{sec:res}
We conduct three experiments. First, we train our models using temporal proximity and compare them with MetaAudio~\cite{metaaudio}, as well as our few-shot inference method using the pretrained feature extractor of CNN14, the same model used for PS. Second, we use PS exclusively for selecting training segments, and we also apply it for both selecting training and testing segments. Thirdly, we perform an ablation study for the augmentation techniques employed in our study.

Table~\ref{Tab:res} shows that SupCon outperforms the meta-learning approaches. In addition, all three SSL methods outperform CNN14 inference, with BT achieving the highest accuracy of 48.90\%. This result underscores the feasibility of learning representations of bird sounds without labeled data. The lower performance of SimCLR and FroSSL in comparison to BT could be explained by the following: When training on long audio files with short events, sampling silence or background noise is possible. If two different samples share this trait in a training batch, the sample-contrastive losses separate them in the embedding space, potentially affecting the quality of the learned representation. 
\begin{table}[h]
\caption{Results of 5-way 1-shot tasks from BirdClef2020}
\label{Tab:res}
\begin{center}
\begin{tabular}{lc}
\hline
\textbf{Method} & \textbf{Accuracy} \\
\hline
\multicolumn{2}{c}{Supervised} \\
\hline
FO-MAML~\cite{metaaudio} & 56.26$\pm$0.45 \\
FO-Meta-Curvature~\cite{metaaudio} & 61.34$\pm$0.46 \\
ProtoNets~\cite{metaaudio} & 56.11$\pm$0.46 \\
SimpleShot~CL2N~\cite{metaaudio} & 57.66$\pm$0.43 \\
Meta-Baseline~\cite{metaaudio} & 57.28$\pm$0.41 \\
SupCon~(ours) & 64.55$\pm$0.42 \\
\hline
\multicolumn{2}{c}{Self-Supervised (ours)} \\
\hline
SimCLR & 43.28$\pm$0.43 \\
FroSSL & 38.24$\pm$0.39 \\
BT & 48.90$\pm$0.42 \\
\hline
CNN14 (PANN) & 32.97$\pm$0.35 \\
\hline
\end{tabular}
\begin{minipage}{7cm}
\centering\fontsize{7}{6}\selectfont Average accuracies along with 95\% CIs using 10,000 test tasks
\end{minipage}
\end{center}
\end{table}

Table~\ref{Tab:ps} illustrates the impact of PS as window selection strategy when applied during training, as well as during both training and test. PS enhances the results of all the methods we experimented with, underscoring the importance of an effective window selection. During inference, PS significantly improves the performance of the nearest prototype classifier by using the windows with the highest PANN activation of the bird class as prototypes, in contrast to averaging all the windows. This approach may be less conservative than averaging, and the expected performance improvement was not evident considering the results obtained with PANN alone (Table~\ref{Tab:res}). The results highlight the complementarity of PS and SSL.
\begin{table}[h]
\caption{Effect of PANN selection of windows}
\label{Tab:ps}
\begin{center}
\begin{tabular}{lc}
\hline
\textbf{Method} & \textbf{Accuracy} \\
\hline
\multicolumn{2}{c}{Supervised} \\
\hline
SupCon & 64.55$\pm$0.42 \\
+PS on Train & 67.68$\pm$0.41 \\
+PS on Train \& Test & 73.07$\pm$0.44 \\
\hline
\multicolumn{2}{c}{Self-Supervised}\\
\hline
SimCLR & 43.28$\pm$0.43 \\
+PS on Train & 55.98$\pm$0.43 \\
+PS on Train \& Test & 64.19$\pm$0.47 \\
\hline
FroSSL & 38.24$\pm$0.39 \\
+PS on Train & 46.23$\pm$0.40 \\
+PS on Train \& Test & 56.79$\pm$0.49 \\
\hline
BT & 48.90$\pm$0.42 \\
+PS on Train & 53.82$\pm$0.42 \\
+PS on Train \& Test & 63.61$\pm$0.47 \\
\hline
\end{tabular}
\end{center}
\end{table}

Table~\ref{Tab:abla} presents the effects of ablating each data augmentation. The results underscore the crucial role played by the combination of both time shifting and spectrogram mixing. This underlines the intuitive idea that a model should be robust to background noise and temporal variation. Together, they contribute significantly to the model's capacity to learn strong representations of bird sounds.
\begin{table}[h]
\caption{Effect of Data Augmentation}
\label{Tab:abla}
\begin{center}
\begin{tabular}{lccccc}
\hline
 & SupCon & SimCLR & FroSSL & BT & RI \\
\hline
 & 73.07 & 64.19 & 56.79 & 63.61 & 24.53 \\
\hline
-SA & 71.20 & 57.85 & 50.14 & 57.73 & \\
-SM & 71.27 & 26.97 & 27.93 & 26.60 & \\
-TS & 72.25 & 58.15 & 27.40 & 27.48 & \\
\hline
\end{tabular}
\begin{minipage}{7cm}
\centering
\fontsize{7}{6}\selectfont \textbf{SA, SM, TS, RI} stand for SpecAugment, Spectrogram Mixing, Time Shift, Random Initialization, respectively.
\end{minipage}
\end{center}
\end{table}
It is important to note that when data augmentation is not enough to learn strong invariances, SimCLR performs the best among the three SSL methods. This may be due to the negative pushing force of its loss, which separates augmented examples coming from different original samples in the embedding space.
\section{Conclusions and Perspectives}
\label{sec:conc}
In this work, we have presented a benchmark of three SSL approaches for few-shot bird sound classification. We have demonstrated that relying mainly on SSL can be competitive with SL. These results using systems relying on the efficient neural network MobileNetV3, provide a promising approach for bioacoustic applications. By employing only time shifting, spectrogram mixing and masking strategies without assuming specific invariances related to bird sounds, we established that meaningful representations can be learned, showcasing promising performance in few-shot learning scenarios. Additionally, leveraging a PANN as relevant segment selection for training and decision at test time significantly improves the performance of classification. 
In future work, we seek to investigate methods for assessing the quality of representations learned through SSL without relying on performance assessment on a validation set, which can substantially differ for the evaluation scenario as in our case.



\bibliographystyle{IEEEbib}
\bibliography{strings,refs}

\end{document}